\documentclass[12pt]{article}

\usepackage{graphicx,amstext,amssymb}
\usepackage{amsmath}

\begin{document}

\title{Characteristic length of dynamical reduction models  and decay of cosmological vacuum
}
\author{S.V. Akkelin\footnote{E-mail: akkelin@bitp.kiev.ua} $^{1}$}
\maketitle
\begin{abstract}
Characteristic length of mass density resolution in dynamical
reduction models  is calculated utilizing  energy conservation law
and viable cosmological model with decreasing energy density of
vacuum (dark energy density). The  value found, $ \sim 10^{-5}$
cm, numerically coincides with  phenomenological spatial
short-length cutoff parameter introduced  in the
Ghirardi-Rimini-Weber model. It seems that  our results  support
the gravity induced mechanism of dynamical reduction.
\end{abstract}

\begin{center}

{\small \textit{$^{1}$ Bogolyubov Institute for Theoretical
Physics, 03680 Kiev-143, Metrologichna 14b, Ukraine \\[0pt]
}}

PACS: {\small \textit{ 03.65.Ta, 03.65.Yz, 42.50.Lc  }}

Keywords: {\small \textit{dynamical reduction models,
characteristic length, dark energy density}}

\end{center}

\section{Introduction}
 The Schr\"{o}dinger equation when applied for
 system that is not isolated  and interacts with
complicated "environment" (it is typical for macro-systems)
results in appearance of the entanglement states involved many
degrees of freedom. The superposition of such states  for all
practical purposes results in the same outcome of measurements as
"mixed" state described by diagonal density matrix. However it
does not mean that superposition disappears, it still exists
globally but is unobservable at either system alone. The
\textit{apparent} decoherence (suppression of interference) for
macro-systems can be explained then as result of the entanglement
of system with its environment (for a review see, e.g.,
\cite{Zeh1}). The density matrix of the corresponding system  in
this approach is just  means for calculating expectation values or
probabilities for outcomes of measurements.

Another approach to decoherence of macro-systems  is grounded on
 idea of stochastic dynamical reduction (for a review see, e.g.,
\cite{Ghirardi1,Pearle}) and treats the decoherence as
\textit{real} process of state vector reduction that takes place
because corresponding  system is influenced by stochastic forces,
in other words, interacts with some \textit{stochastic}
environment.  The nature of the stochastic field is usually not
specified there, except for the dynamical reduction models
involving gravity, so called Newtonian Quantum Gravity approach
\cite{D-L,D1,D2,Ghirardi2,Penrose}, where stochastic field is
associated with non-relativistic gravity potential field. Reasons
why gravity can be treated as stochastic field at small space-time
scales are known for a long time (see Ref. \cite{Bronstein}):
inasmuch as refined length measurement requires large momentum, a
probe itself disturbs the gravity field curving space-time and
distorting the interval one seeks to measure leading to limit of
measurability of gravity field (metric tensor). In fact, it is
consequence of the equivalence principle.

Noteworthy that spatial  reduction of wave function by stochastic
forces leads to  momentum increase because of the uncertainty
principle, and, therefore, to increase of particle energy. It
results in apparent violation of the conservation laws, and the
rate of increase diverges for a point-like object
\cite{Ghirardi2}. To alleviate the problem  the spatial coarse
graining of the mass density (cutoff on spatial mass density
resolution) is  used in the dynamical reduction models, this
cutoff parameter fixes also the distance scale (localization
width) beyond which the wave function collapse becomes effective.
The value of the short-length cutoff, $ 10^{-5}$ cm, that was
proposed in Ref. \cite{Ghirardi} (see also \cite{Ghirardi2}),
keeps energy non conservation rate below experimental limits.
Certainly, to explicitly satisfy the conservation laws one needs
account for not only particle contribution, but also contribution
of the stochastic field to the conserved quantities.

 Because, as we discussed above, the most natural candidate
for stochastic field of the dynamical reduction models is gravity,
and a value of the scale factor governs the rate of energy
transfer from stochastic field to the matter particles, it might
be expected that a value of the scale factor is related to the
rate of decay of cosmological (gravitational) vacuum and,
consequently, related to decrease  of the vacuum energy density
with time. Note that cosmological scenarios of the universe
evolution where $\Lambda(t)$ decreases slowly with  time
($\Lambda$ represents the energy density of the vacuum) are
allowed by the observational cosmology (see, e.g.,
\cite{Lima,Alcaniz1}).

 In this paper we utilize the correlation function of fluctuations
  of gravity acceleration field \cite{D-L} (see also \cite{D1}) and
  viable model of the cosmological vacuum decay
\cite{Alcaniz} to estimate value of the spatial short-length
cutoff parameter of the dynamical reduction models.\footnote{The
relation of the particle energy increase  with the loss of vacuum
gravitational energy has been also discussed in Refs.
\cite{Pearle,Pearle1}.} In  Section $2$ the cosmological model
\cite{Alcaniz} with decaying vacuum energy density  is briefly
reviewed. The rate of the cosmological vacuum decay found in the
model is utilized in  Section $3$ to calculate the characteristic
length of the dynamical reduction models involving gravity.  In
Section $4$ conclusions are given  and some possible consequences
for biogenesis are briefly discussed.

\section{Cosmological model with time-varying cosmological constant and  decay of vacuum}
Let us first briefly review  the cosmological  vacuum decay
scenario recently proposed in the Ref. \cite{Alcaniz} (see also
\cite{Alcaniz1}). We choose this model among many others
phenomenological $\Lambda (t)$ models (see, e.g.,
\cite{Overduin,Lima,Alcaniz1}) because it was explicitly
demonstrated in Refs. \cite{Alcaniz,Alcaniz1} that this model is
compatible with current cosmological data.

 The Einstein field equations  are (throughout the paper we  use  the
"natural units" where $\hbar=c=1$)
\begin{eqnarray}
R^{\mu \nu} - \frac{1}{2} g^{\mu \nu} R = 8 \pi G \left [T^{\mu
\nu} + \frac{\Lambda}{8 \pi G} g^{\mu \nu} \right ],
 \label{v-1}
\end{eqnarray}
where $T^{\mu \nu}$ is the energy-momentum tensor of "ordinary"
(nearly $3 \%$ of total energy density) and "dark" (nearly $ 27
\%$) matter, while  $\Lambda$ is the cosmological constant that is
responsible for famous "dark" energy density (nearly $70 \%$)
contribution to total energy density,  and also for acceleration
of the universe expansion at the present epoch. The cosmological
constant can be treated as the energy density of the cosmological
vacuum, $\epsilon_{vac}$. Since a vacuum has equation of state
$\epsilon_{vac}=-p_{vac}$, $p$ is pressure, then (for recent
review see, e.g., \cite{Pad1})
\begin{eqnarray}
\frac{\Lambda}{8 \pi G} g^{\mu \nu}=T^{\mu
\nu}_{vac}=\epsilon_{vac}g^{\mu \nu}.
 \label{v-1-1}
\end{eqnarray}
 The nature of "dark" matter is still unclear, most probably (see, e.g.,
\cite{Majumdar})  some weakly interacting particles are
responsible for the "dark" matter energy density contribution to
the energy-momentum tensor of matter fields.

According to Bianchi identities
\begin{eqnarray}
\partial_{\nu} (R^{\mu \nu} - \frac{1}{2} g^{\mu \nu} R )=0,
 \label{v-2}
\end{eqnarray}
 therefore if $\Lambda$ depends on
time (i.e. vacuum decays in the course of the expansion) then
 $T^{\mu \nu}$ can not be separately conserved and there is a
 coupling between  $T^{\mu \nu}$ and $\Lambda$ that follows from
 Eq. (\ref{v-1}):
\begin{eqnarray}
u_{\mu} \partial_{\nu} T^{\mu \nu}= - u_{\mu} \partial_{\nu} \left
( \frac{\Lambda}{8 \pi G} g^{\mu \nu} \right ),
 \label{v-3}
\end{eqnarray}
here $u_{\mu}(x)$ is local 4-velocity of cosmological expansion.
The assumption of isotropy and homogeneity implies that the large
scale geometry can be described by a metric of the form
\begin{eqnarray}
ds^{2}=dt^{2}-a^{2}(t)d\textbf{r}^2,
 \label{v-3-g}
\end{eqnarray}
 here $a(t)$ is scale
(expansion) factor of the universe and   flat spatial sections are
also  assumed. Taking  then perfect fluid form for $T^{\mu \nu}$,
\begin{eqnarray}
T^{\mu \nu}=(\epsilon + p )u^{\mu}u^{\nu} - pg^{\mu \nu},
 \label{v-4}
\end{eqnarray}
one can get
\begin{eqnarray}
\dot{\epsilon}+(\epsilon + p)\partial_{\alpha}u^{\alpha}=-
\frac{\dot{\Lambda}}{8 \pi G},
 \label{v-5}
\end{eqnarray}
where the overdot denotes covariant derivative along the world
lines (time comoving derivative, for instance,
$\dot{\epsilon}=u_{\alpha}\partial^{\alpha}\epsilon$). For
homogeneous and isotropic  Friedmann-Robertson-Walker geometry
\begin{eqnarray}
\partial_{\alpha}u^{\alpha}=3 H,
 \label{v-6}
\end{eqnarray}
where
\begin{eqnarray}
 H(t) = \frac{\dot{a}}{a}
  \label{v-6-h}
\end{eqnarray}
is the Hubble parameter, it measures the rate of expansion of the
universe.

Then, if vacuum transfers energy to matter, the question arises
where the matter stores the energy received from the vacuum decay
process. The traditional approach for  the vacuum decay process is
vacuum decay into matter particles. Following to Ref.
\cite{Alcaniz}, one can assume that vacuum decay results (mainly)
in creation of weakly interacting "dark" matter particles and, so,
neglect a contribution of the "ordinary" matter to the left hand
side of Eq. (\ref{v-5}). It allows  to relate the time dependence
of the energy density of the vacuum with temporal evolution of
the "dark" matter. Then, assuming that "dark" matter is
pressureless, $p_{d}=0$ (cold "dark" matter model), one can  get
the following equation:
\begin{eqnarray}
\dot{\epsilon}_{d}+3\frac{\dot{a}}{a}\epsilon_{d}\approx
-\dot{\epsilon}_{vac}.
 \label{v-7}
\end{eqnarray}

Because vacuum decays into "dark" matter the latter  will dilute
more slowly compared to its standard evolution, $\epsilon_{d}\sim
a^{-3}$,  when  the vacuum energy density does not change in the
course of the expansion, $\dot{\epsilon}_{vac}=0$. Then making a
specific  ansatz for the "dark" matter energy density
\begin{eqnarray}
\epsilon_{d} = \epsilon_{d0}\frac{a^{3-\delta}_0}{a^{3-\delta}},
 \label{v-8}
\end{eqnarray}
where $\delta>0$ is a constant that characterizes the deviation
from the standard evolution, $\epsilon_{d0}=\epsilon_{d}(t_{0})$
and $a_0=a(t_{0})$ are the current values of the "dark" matter
energy density and of the scale factor of the universe
respectively, we have
\begin{eqnarray}
\epsilon_{vac} = \widetilde{\epsilon}_{vac} +
\frac{\delta}{3-\delta}\frac{a^{3-\delta}_0}{a^{3-\delta}}\epsilon_{d0}
 \label{v-8}
\end{eqnarray}
where $\widetilde{\epsilon}_{vac}$ does not depend on time.

It was found from analysis of cosmological data  that $\delta =
0.06 \pm 0.10$  \cite{Alcaniz1}. Such a relatively slow decrease
of $\Lambda (t)$ (vacuum energy density) indicates, perhaps, that
viable cosmological variable-$\Lambda$ models with cold "dark"
matter (CDM) and standard physics  should not differ too
drastically from concordance $\Lambda$CDM model to be compatible
with observational cosmology. In the next Section we study whether
the found in the  Refs. \cite{Alcaniz,Alcaniz1} rate of
cosmological vacuum decay  is compatible with the rate of particle
energy increase presupposed  in the dynamical reduction models,
and, so, whether the corresponding characteristic length  can be
defined by $\Lambda(t)$CDM cosmology.

\section{Characteristic length
for given rate of the cosmological vacuum decay}

To perform the corresponding analysis let us assume that, while
main  energy gain from cosmological vacuum decay is adopted by the
"dark" matter, the cosmological vacuum couples not only to the
"dark" matter but to the "ordinary" matter as well, and assume
that vacuum decay does not lead to creation of  "ordinary" matter
particles but increases the mean kinetic energy of the ones. Then,
if corresponding kinetic energy gain attributed to all the
particles in the universe is much smaller than the total loss of
vacuum energy in the   universe,   it can not essentially
influence on the results of Ref. \cite{Alcaniz} that are briefly
reviewed in the previous Section.  The next step then is to relate
the energy gain of a particle, whose mass is locally smeared
within the corresponding characteristic volume, with the rate of
energy loss of vacuum within the same volume. It gives us, in
fact, upper limit to the particle energy increase allowing by the
loss of vacuum energy.

 For the sake of simplicity, let us
consider a energy gain of a single nucleon. Let us assume that
decay of the cosmological vacuum induces stochastic gravity field
that results in stochastic acceleration of a particle. Then the
mean non-relativistic kinetic energy induced by the vacuum decay
is
\begin{eqnarray}
E_{part}(t)= \frac{m}{2} \langle \mathbf{v}^{2}(\mathbf{r},t)
\rangle =
\frac{m}{2}  \stackrel{t}{%
\mathrel{\mathop{\int }\limits_{t_{i}}}%
} \stackrel{t}{%
\mathrel{\mathop{\int }\limits_{t_{i}}}%
} \langle \mathbf{g}(\mathbf{r},t^{\prime})
\mathbf{g}(\mathbf{r},t^{\prime \prime})\rangle dt^{\prime}
dt^{\prime \prime},
 \label{s-3-1}
\end{eqnarray}
where $\textbf{g}$ and $\textbf{v}$ are  stochastic acceleration
and velocity field respectively, $ \langle ...\rangle $ means the
averaging over the corresponding characteristic volume $V_{c}$,
and we assume that initially $\textbf{v}$ is equal to zero,
$\textbf{v}(t_{i})=0$. To proceed we need in  correlation function
of fluctuations of  acceleration field, $\mathbf{g}$.  The
analysis of measurability of the Newtonian acceleration field that
was done in Ref. \cite{D-L} (see also \cite{D1}) results in the
following expression for correlation function (assuming that
$\langle \mathbf{g}\rangle = 0$):
\begin{eqnarray}
\langle \mathbf{g}(\mathbf{r},t^{\prime})
\mathbf{g}(\mathbf{r},t^{\prime \prime})\rangle \sim
\frac{G}{V_{c}} \delta (t^{\prime }-t^{\prime \prime}).
\label{c-1-1}
\end{eqnarray}
Hereafter we will take Eq. (\ref{c-1-1}) as equality. Hence we
obtain from Eqs. (\ref{s-3-1}) and (\ref{c-1-1})  that
\begin{eqnarray}
 \langle \mathbf{v}^{2}(\mathbf{r},t)
\rangle  = G \stackrel{t}{%
\mathrel{\mathop{\int }\limits_{t_{i}}}%
} \frac{dt^{\prime}}{V_{c}(t^{\prime})}, \label{c-1-2}
\end{eqnarray}
and then the rate of the particle energy gain,
$\frac{dE_{part}}{dt}$, is
\begin{eqnarray}
\frac{dE_{part}}{dt}= \frac{mG}{2V_{c}(t)}.
 \label{s-3-2-E}
\end{eqnarray}
Taking into account the energy conservation, we get finally the
equation
\begin{eqnarray}
 \frac{mG}{2V_{c}(t)} = - V_{c}(t) \dot{\epsilon}_{vac}(t)
 \label{s-3-2-1}
\end{eqnarray}
and, therefore,
\begin{eqnarray}
 V_{c}(t)= \left (-\frac{mG}{2\dot{\epsilon}_{vac}(t)} \right
 )^{1/2}.
 \label{s-3-2-v}
\end{eqnarray}
 One can see from
Eq. (\ref{v-8}) that
\begin{eqnarray}
\dot{\epsilon}_{vac}= - \delta \cdot \epsilon_{d0} \cdot
\frac{\dot{a}}{a} \cdot \frac{a^{3-\delta}_0}{a^{3-\delta}}.
 \label{s-1}
\end{eqnarray}
Then one can conclude from the above expressions  that $V_{c}(t)$
increases (in cosmological sense, i.e. very slowly)  with time.

 Now let us estimate  the present ($t=t_0$) value of the characteristic volume, $V_{c}(t_{0})$.
Taking into account that
\begin{eqnarray}
 \dot{\epsilon}_{vac}(t_0)=
- \delta \cdot H_{0}\cdot \epsilon_{d0},
 \label{s-2}
\end{eqnarray}
where $H_{0}$ is current value of the Hubble parameter,
$H_{0}=H(t_{0})$,  we get for the characteristic volume
 \begin{eqnarray}
V_{c}(t_{0})=\left (\frac{mG}{2\delta  H_{0} \epsilon_{d0}} \right
)^{1/2}
 \label{s-6}
\end{eqnarray}
and, therefore,  for the characteristic length   $R_{c}\equiv
V_{c}^{1/3}$
 \begin{eqnarray}
R_{c}(t_{0})= \left (\frac{m \cdot G}{2  \delta  H_{0}
\epsilon_{d0}}\right )^{1/6}.
 \label{s-7}
\end{eqnarray}

The total energy density of the universe is close to the critical
energy density (see, e.g. \cite{Pad1}),
$\epsilon_{crit}(t_{0})=\frac{3 H_{0}^{2}}{8 \pi G}$. Then,
because "dark" matter energy density is approximately equal to
$27\%$  of the total energy density, we get
\begin{eqnarray}
\epsilon_{d0}=0.27 \cdot \frac{3 H_{0}^{2}}{8 \pi G}
 \label{s-8}
\end{eqnarray}
and, finally,
 \begin{eqnarray}
R_{c}(t_{0})= \left (\frac{8 \pi \cdot m \cdot G^{2}}{6 \cdot 0.27
\cdot \delta \cdot H_{0}^{3}}\right )^{1/6}.
 \label{s-9}
 \end{eqnarray}
 Here  $\delta = 0.06$, $H_{0}
=\frac{1}{1.3 \cdot 10^{28}}$ cm$^{-1}= 0.769 \cdot 10^{-42}$ GeV,
$G=1/M_{Pl}^{2}$, $M_{Pl}=1.22 \cdot 10^{19}$ GeV, and $m$ is
assumed to be proton mass, $m=0.938$ GeV. Also taking into account
 that $1$ Gev$^{-1}$= $1.973 \cdot 10^{-14}$ cm we get finally
$R_{c}(t_{0})=1.06 \cdot 10^{-5}$ cm.  Then the characteristic
volume multiplied by the number of particles making up
 "dark" and "ordinary" matter is much less than the volume of the
 universe, justifying thereby  neglect of influence of the vacuum energy
 transfer to the kinetic energy of particles on temporal evolution
 of the vacuum energy density.

Interestingly enough that practically the same value, $10^{-5}$
cm, was proposed in Ref. \cite{Ghirardi} for phenomenological
spatial cutoff parameter of Quantum Mechanics with Spontaneous
Localizations, and is  accepted now \cite{Ghirardi1,Adler} as low
limit for the characteristic  length in the dynamical reduction
models. Then our finding, perhaps,  supports the gravity induced
mechanism of dynamical reduction.

Note that $m$-dependence of $R_{c}$  is rather weak, $R_{c} \sim
m^{1/6}$, and therefore $R_{c}$ does not change much even for
electron mass. Also this $m$-dependence does not spoil the fast
decoherence of the distant, $r \gg R_{c}$, superposition of
macro-objects, because then decoherence time, $t_{dec}$, is
$t_{dec} \sim R_{c}/G m^{2}$ \cite{D2} and, consequently, $t_{dec}
\sim m^{-11/6}$ resulting, as well as  for $m$-independent
$R_{c}$, in extremely small value  that is desired property for
Schr\"{o}dinger cat states.

\section{Concluding remarks}
  Summarizing this work, we point out that apparent
violation of the energy conservation  is not actually  shortcoming
of the dynamical reduction models.  Moreover, energy conservation
law  gives deep insight into the physics of the spatial cutoff
parameter.   Namely, we found that the value, $ \sim 10^{-5}$ cm,
of the characteristic length is, perhaps,  conditioned by the
present rate of energy transfer from decaying vacuum to a
particle. Because a value of the rate of the vacuum decay need be
compatible with the observational cosmology, one can say that, in
a certain sense, cosmology dictates the characteristics of
dynamical reduction.

One can speculate that increase of $V_{c}$ with cosmological time
(see Eq. (\ref{s-3-2-v})) is related to the problem of appearance
of life and consciousness. First of all, note that biogenesis on
Earth has occurred very rapidly, during $0.1^{+0.5}_{-0.1}$ Gyr
and, therefore, appearance of life could be highly probable
process \cite{Davis}. However, while age of Earth, $\approx 4.566$
Gyr, is much lower than age of the universe, $\approx 13.7$ Gyr,
we do not observe signals of a vital activity of extraterrestrial
civilizations ("silence of the universe" problem, see, e.g.,
\cite{Olum}). Explanation of the apparent paradox can be the
following. If quantum mechanics indeed play a key role in the
origin of biological organisms\footnote{For instance, since
quantum system can exist in superposition of states, searches of
biologically potent molecular configurations may proceed much
faster than one might expect using classical estimates, leading,
thereby,  to the fast biogenesis (see, e.g., recent reviews
\cite{Davies} and references therein).}, then spatial cutoff
parameter value (that regulates the "borderline" between quantum
and classical worlds) is extremely important for origin and
operation of biological organisms. Therefore an appropriate value
of the cutoff parameter $R_{c}$ fixes the cosmological time when
appearance of living organisms becomes possible. Noteworthy that
formation of Earth   fell roughly at the same time interval when
expansion of the universe became accelerated \cite{start}. So, one
can speculate that necessary conditions for origin of a life
appear only recently, during the epoch of accelerated expansion
when the characteristic length also increases with acceleration.
Then the observational lack of extraterrestrial intelligent life
can be consequence of the "time cutoff" for the most early
emergence of a life in the universe.

\end{document}